# Is an historical economic crisis upcoming?

Çağlar Tuncay
caglart@metu.edu.tr

**Abstract:**
In this work, the time chart of Dow Jones Industrial Average (DJIA) index is analyzed and approach of recession time term is predicted, which may be hallmark of a worldwide economic crisis. However, the methods used for the prediction will be disclosed a few years from now. On the other hand, this work will be updated by the author frequently once in every few months where no revisions will be made on the previous uploads and a timestamp will designate each part. Thus, the time evolution of the crisis can be followed and the prediction may be verified by the readers in time.

**Introduction:**
Various social phenomena such as economy are known (or believed) to be non-deterministic, and hence unpredictable. Yet, many approaches are available in the literature to investigate the time series of numerous assets or stock exchange indices from the world market, or prices of valuable materials such as gold or petroleum. One may simply enter the keywords for econophysics, exchange, asset, gold, petroleum in the search box of favorite search engines (you can even use the one on the upper right corner of this webpage) and download the relevant papers.
The author of this work, whose works may also be accessed and investigated following the way described in the previous paragraph, developed several new methods for the prediction of the future of prices or stock exchange indices without publishing them, which should be treated apart from the published or publicized ones.

**The index of DJIA:**
The daily time chart of the closes for the DJIA index ($\chi(t)$) with a time domain from the establishment up to 30 May 2010 is depicted in Fig.1 where the vertical axis is logarithmic. On inspection of Fig.1 it may be observed that the overall behavior can be investigated in five epochs.

EPOCH 1: This epoch begins with the establishment of the index at 01 Dec 1928 and lasts till about 01 Jan 1943 ($\chi=120.25$). Hence, duration of the epoch is nearly 3,500 trading days which covers the Great Depression. The index metrics show several (nearly) exponential increasing and decreasing behaviors which appear as (nearly) straight lines in Fig.1:

$$\chi(t) \propto \exp(\alpha t) \qquad (1)$$

where the exponent $\alpha$ is different for different time intervals in that epoch.
It should be noted that the amplitude of fluctuation in the exponential behavior decreases with time where two straight lines function as the lower and upper bounds; the arrows $A_1$ and $A_2$ in Fig.1, respectively. The upper (lower) bound line passes through the locally maximum (minimum) index values. Moreover, when the values of an index (or prices of an asset) are bounded by two lines which approach to each other with time then the metrics are said to be squeezed in chartist language. The reader is referred to (Schmidt, 2002) for a discussion on chartist approach (technical trader) and efficient market theory.

The log-linear slopes of the bounding lines can be estimated using Eq. (1). For example, that slope, or the exponent α in Eq. (1), for the supporting line can be obtained approximately as follows:

$$\alpha = (\ln(\chi(t_2)) - \ln(\chi(t_1)))/(t_2 - t_1) \qquad (2)$$

since $\ln(\exp(\alpha t))$ is known to be $\alpha t$. If the given dates (for time t) and the corresponding index values ($t_1$=946, $t_2$ = 3,380 days, and $\chi(t_1)$=92.92 and $\chi(t_2)$= 41.22) are used in Eq. (2) then the result comes out approximately as 0.0003 per day.

EPOCH 2: The second epoch begins at around 01 Jan 1943 and continues till nearly 13 Feb 1964 (the 8854th day with χ=794.42). The duration of this epoch is approximately 5,300 trading days and exponential growth is the main characteristic.

The exponent (α, in Eq. (1)) of the exponential growth during the current epoch can be estimated in terms of Eq. (2). If the relevant dates, 01 Jan 1943 ($t_1$=8,854) and 13 Feb 1964 ($t_2$=14,148), and the corresponding index values $\chi_1$=120.25 and $\chi_2$=794.42, respectively are substituted in Eq. (2) then the result can be obtained approximately as 0.0003 per day. Hence, the behavior of the DJIA index during the current epoch may be treated in the first order approximation; i.e., omitting the decorations on the time profile, as an exponential growth with the exponent 0.0003 per day in linear scale.

It should be noted that the slope of the supporting line (lower bound) in the first epoch (the arrow $A_1$ in Fig. 1) is also equal to 0.0003 per day. This means that the supporting lines of the first and second epochs are the same.

EPOCH 3: The third epoch begins at around 13 Feb 1964 and continues till around 01 Jan 1982 (13,340th day with χ=1000); hence, the duration of this epoch is around 4,488 trading days or 20 calendar years. The characteristic of this epoch is the horizontal behavior of the index in a channel between nearly 800 and 1000; see the arrows $A_4$ and $A_5$ in Fig. 1, respectively. In other words, the lines for the upper and lower bounds are horizontal; i.e., α = 0 for both lines, in this epoch. The reader may be referred to (Dempster and Jones, 2002) for the technical analyses of the prize channels.

It should be noted that the metrics fall below the supporting line of the previous epochs at a day around 17 June 1989 and does not crosses it, later.

EPOCH 4: This epoch continues till the beginning of 2000 crash, around the 18000th trading day and hence the duration of this epoch is around 4,660 days. Actually, this epoch consists of two time terms each displaying exponential growth with different exponents of 0.0005 per day between 01.01.1982 and 08.12.1994, and 0.0008 per day as designated by the arrows $A_6$ and $A_7$ in Fig. 1, respectively.

EPOCH 5: Fluctuations about the level of 10,000 is the main characteristics of the current epoch. Hence, horizontal behavior is expected to continue for about 20 calendar years from now.

Determination of the upper and lower bounds of the channel for the last epoch is not possible since the epoch is not terminated. However, it may be expected that the upper bounding line will pass through the value of 14,160 which is attained by the index at 09 Dec 2007; see $A_8$ in Fig. 2.

**Discussions:**

It is claimed that two behaviors are dominant in the time course of the DJIA index; horizontal fluctuations around the index values 100, 1,000 and 10,000, and exponential growths in-between. Moreover, the news, economical or political, say are not decisive on the topography of the DJIA index. The reader is referred to (Joulin et al., 2010) for a discussion about the dependence of the stock prices on news and volume.

**Predictions and Conclusions:**

The following predictions are made relying on the unpublished methods:
1. The last and upcoming few months ahead, which are obviously negative with respect to the last few years, are the best times with respect to the several months in the future. To be more precise, the upcoming several weeks or months will be worse and worse with time. For example, the DJIA index will test 6000's within the coming 10 to 12 months; i.e. around July 2011.
2. It is expected that the fall in the stock exchanges will continue till DJIA tests 4000's at a time around September 2011.

This author hopes that he is wrong in all the mentioned considerations, and the above negativities remain as nothing more than nightmares.

It is worth nothing that the following remark, which is found in (Tuncay and Stauffer, 2007) may be interesting for the reader: "However, the fluctuations in the financial value V(t) were quite strong in 1940 at the end of the great recession, and also in recent months. Perhaps these recent fluctuations signal a transition to a different regime, to be seen in the coming years." Moreover, the over-all discussion and conclusion will be given in the final part of this series.

**FIGURES**

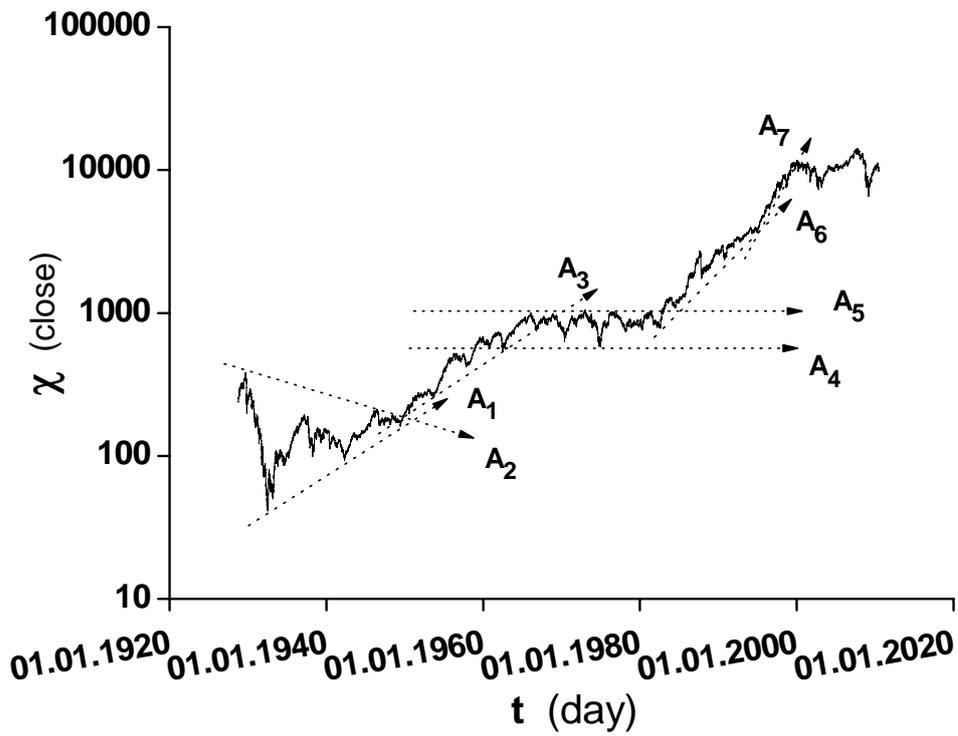

**Figure 1**   The time profile of the index of DJIA from the establishment till 30 May 2010.

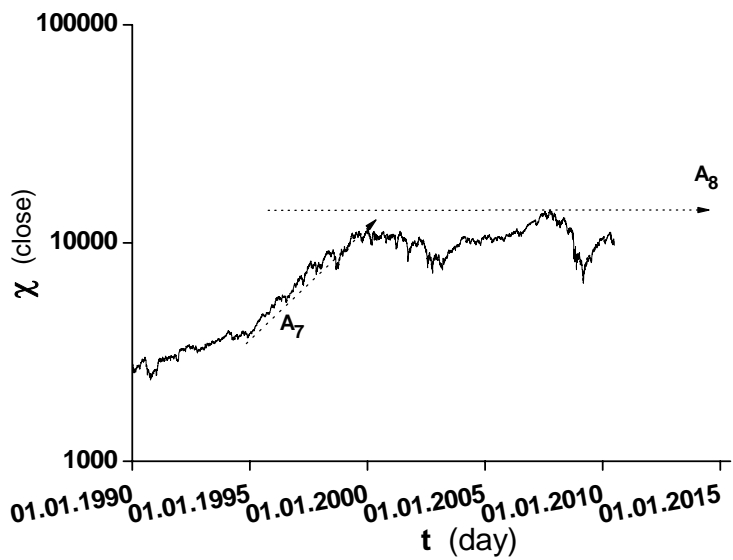

**Figure 2**   The same as Fig.1 but with a different time and values domain.